\documentclass[twocolumn,showpacs,preprintnumbers,amsmath]{revtex4}
\usepackage{dsfont,epsfig,amsmath,amssymb,rotating}
\usepackage{graphicx}

\begin{document}

\title{Dynamics of Enzyme Digestion of a Single Elastic Fiber Under Tension: An Anisotropic Diffusion Model}

\author{Asc\^anio D. Ara\'ujo$^{1,2}$, Arnab Majumdar$^{1}$, Harikrishnan Parameswaran$^{1}$ and B\'ela Suki$^{1}$}

\affiliation{$1$ Department of Biomedical Engineering, Boston University, Boston, Massachusetts\\
 $2$ Departamento de F\'{\i}sica, Universidade Federal do Cear\'a, Fortaleza, Cear\'a, Brazil.}

\date{\today}

\begin{abstract}
We study the enzymatic degradation of an elastic fiber under tension using an anisotropic random-walk model, coupled with binding-unbinding reactions that weaken the fiber. The fiber is represented by a chain of elastic springs in series, surrounded by two layers of sites along which enzyme molecules can diffuse. Through numerical simulations we show that the fiber stiffness decreases exponentially with two distinct regimes. The time constant associated with the first regime decreases with increasing applied force, which is in agreement with published experimental data. In addition, a simple mean field calculation allows us to partition the time constant into geometrical, chemical and externally controllable factors, which is corroborated by the simulations.
\end{abstract}

\maketitle

The extracellular matrix (ECM), the biological structure that supports cells, is composed of elastic fibers such as elastin and collagen. The complex organization of these fibers undergoes a continuous maintenance that requires the catalytic action of enzymes, called proteases \cite{Ghajar_08}. In diseases, such as pulmonary emphysema, tissue destruction is thought to be a consequence of the imbalance between protease and antiprotease activities leading to degradation of elastin fibers \cite{Barnes_00}. Biological tissues in vivo are also under tension which may interfere with the enzymatic activity. Indeed, recent experiments show that mechanical stretch accelerates the rate of degradation of engineered ECM during elastase-induced digestion \cite{Rajiv_07}.

The elasticity of a single fiber depends on how its molecular constituents are organized. During digestion, the molecules in the fiber as well as the cross-links can be cleaved by enzymes causing fiber stiffness to decreases. Furthermore, following cleavage, an enzyme can unbind, diffuse, bind at a different location and cleave another molecule. This leads to the question: How are the diffusion and binding of the enzyme and the subsequent degradation of the fiber affected by the presence of an external mechanical force on the fiber?

In this letter, we study the decay of stiffness of a single fiber under tension during enzymatic digestion using an anisotropic random walk model coupled with binding-unbinding reactions. To our knowledge this is the first investigation of the mechanical properties of a single fiber that takes into account the simultaneous effects of enzyme diffusion, binding, cleaving and mechanical force.

Several different diffusion-reaction models have been used to describe processes at the level of ECM, cell membranes, macromolecules and DNA \cite{Abete_04,Dahirel_09,Schmit_09,Berry_02,Michael_05,Gennes_82}. In addition, the random walk is often used as a diffusion model that takes into account the morphological details of the system. Spring network models also provide a useful framework for analyzing the changes in the mechanical properties of ECM sheets \cite{Kantor_84,Bouchiat_98}. Here we use a modified random walk to mimic anisotropic diffusion of enzyme particles along a fiber and study the digestion of the fiber represented by a chain of springs.

\begin{figure}
\begin{center}
\includegraphics[width=8cm]{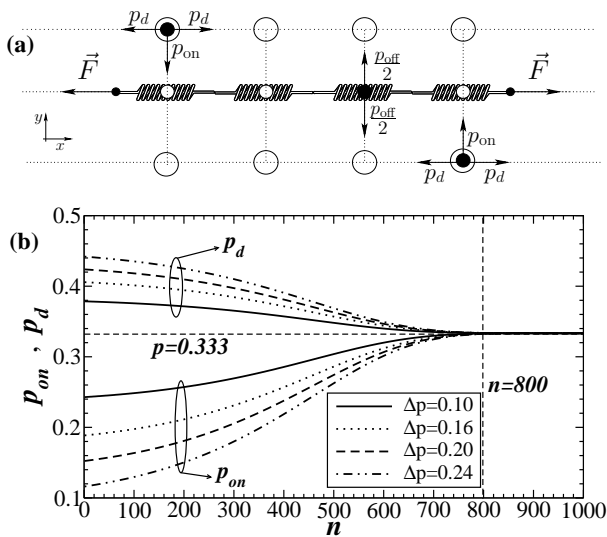}
\caption{$(a)$ Schematic diagram of the chain of springs and sites used in our model. The binding sites on the springs and the two layers of sites are represented by small and big open circles, respectively. The set of enzyme particles are shown as filled black circles. The particle at the botton layer can move up, left, or right while the particle at the top can move down, left, or right. The particle on the spring can move only up or down. $(b)$ The binding probability $p_{\text {on}}$ defined by Eq.~(\ref{eq_1}) and the diffusion probability $p_{d}$ as a function of the number of visits $n$ at a fixed site. Lines of different styles correspond to $\Delta p=0.10$ (solid lines), $\Delta p=0.16$ (dotted lines), $\Delta p=0.20$ (dashed lines) and $\Delta p=0.24$ (dash-dotted lines). The lines above and below the horizontal dashed line at $\Delta p=0.333$ corespond to $p_{d}$ and $p_{\text {on}}$, respectively. The vertical dashed line represents the region where the isotropic behavior with $p=0.333$ is reached.}
\label{fig_1}
\end{center}
\end{figure}

Our model consists of a one-dimensional chain of $N_{s}$ linearly elastic springs in series representing an elastic fiber as in Fig.~\ref{fig_1}$a$. The fiber is surrounded by two layers of sites along which particles representing enzymes can diffuse. Periodic boundary conditions are applied in the $x$ direction. Both ends of the chain are subjected to a constant force $\boldmath{F}$ during digestion which mimics tension in the fiber.

In order to simulate enzyme activity on the fiber, we begin with a chain having identical initial spring constants $k(t=0)\equiv k_{0}$. The diffusion of enzyme is initiated by releasing a set of particles at random positions in the two layers. Each particle moves according to a set of probabilistic rules, controlling the diffusion and reaction processes: (i) $p_{d}$ is the probability for a particle to move right or left, parallel to the chain. This probability is associated with diffusion. (ii) $p_{\text {on}}$ is the probability for a particle to move up from the bottom or down from the top layer. This step represents an enzyme molecule binding to a binding site on the fiber. Only one particle can be bound to a single spring at any time. (iii) $p_{\text {off}}$ is the probability for a bound particle to move up or down to the top or bottom layer, respectively. This step is related to an enzyme molecule unbinding from the fiber. Once a bound particle unbinds, the local spring contant $k$ is reduced by a constant factor $\gamma$, $k\rightarrow\gamma k$. The probabilities $p_{\text {on}}$ and $p_{d}$ are related by the constraint $p_{\text {on}}+2p_{d} =1$.

Local anisotropy is introduced through the probability $p_{\text {on}}$ which depends on the local spring constant $k$,
\begin{equation}
 p_{\text {on}}=\frac{1}{3}-\Delta p e^{-\lambda(F/k)}
\label{eq_1}
\end{equation}
where $\Delta p$ is the initial anisotropy and $\lambda$ is a characteristic length. Eq.~(\ref{eq_1}) expresses the fact that when the enzyme cleaves the fiber, the local $k$ decreases by a small amount. Since $\boldmath{F}$ is constant, the local stretch increases and either more binding sites appear or the binding of an enzyme becomes easier which we represent by an increase in the local $p_{\text {on}}$. This introduces anisotropy in the particle movement and makes the enzyme activity dependent on the local $k$. The parameter $\Delta p$ is related to difficulty of an enzyme molecule to reach a binding site which depends on the surface roughness of the fiber. In Fig.~\ref{fig_1}$b$ we can see that at the beginning of the diffusion, the number of times a spring has been visited by particles is $n\approx 1$ and the initial values of both $p_{\text {on}}$ and $p_{d}$ depend on $\Delta p$. As the diffusion progresses, $p_{\text {on}}$ increases slowly. Around $n=100$, $p_{\text {on}}$ increases significantly until it approaches the isotropic value case $p_{\text {on}}=p_{d}=1/3$, where the rate of increase is reduced. When $n\approx 800$, the diffusion reaches the isotropic regime in which both diffusion and binding are equally likely. We assume that in this regime, locally the fiber is at its unfolding limit in that the number of binding sites remains constant.

\begin{figure}[b]
\begin{center}
\vspace{1.0cm}
\includegraphics[width=8cm]{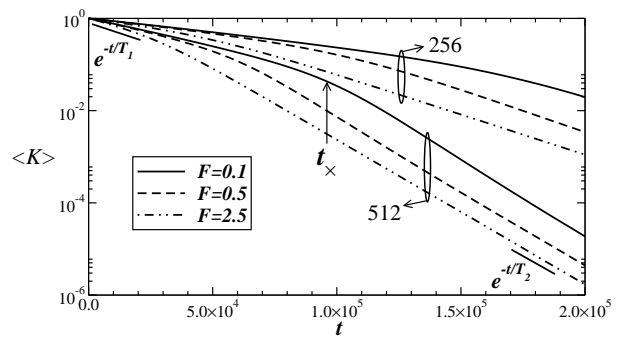}
\caption{Log-linear plot of the averaged stiffness $\langle K \rangle$ as a function of diffusion time $t$ for different values of $F=0.1,0.5,2.5$ and $N_{p}=256,512$ with $\Delta p=0.20$ and $p_{\text {off}}=0.5$. The black solid line segments at the beginning and end of the simulations represent exponential fits to estimate the value of the time constants $T_{1}$ and $T_{2}$, respectively.}
\label{fig_2}
\end{center}
\end{figure}

The probability $p_{\text {off}}$ is determined by the molecular properties of the specific enzyme and its substrate and is the same for each spring, for all times. A particle may remain bound for more than one time step, with probability $1-p_{\text {off}}$. We assume that cleavage occurs during unbinding so that $k$ is reduced only when the particle unbinds from the spring. Thus, $k$ for each spring is a function of time because it decreases as the spring is repeatedly visited by particles. Since $k$ decreases by the constant factor $\gamma$, we can write the relation $k(t)=\gamma^{n}k_{0}$, where $n$ is the total number of visits by time $t$. The stiffness $K(t)$ of the fiber is calculated as the equivalent stiffness of all $N_{s}$ springs connected in series, $K(t)=1/\sum_{i=1}^{N_{s}}[k_{i}(t)]^{-1}$.

Next, we study the evolution of $K(t)$ for different sets of parameters. We use a chain composed of $N_{s}=10^{4}$ springs, different numbers of particles $N_p=256,512,1024$ and different values of the external force $F$ within the interval $[0.1,2.5]$. The $N_{p}$ and $F$ are related to the experimentally controllable macroscopic parameters of the digestion process. At the microscopic level, we vary $p_{\text {off}}$ between $[0.1,1.0]$ while $\Delta p$ is chosen from the interval $[0.10,0.24]$. Additionally, there are three constant parameters: $\lambda=0.10$, $\gamma=0.995$ and $k_{0}=1$. We obtain the time course of $K(t)$ for $t=2 \times 10^{5}$ time steps, where at each time step, we attempt to move all $N_{p}$ particles in the system. We repeat the digestion simulation $500$ times with different realizations and average $K(t)$ over all runs.

The results for $\langle K(t) \rangle$ are plotted on a log-linear scale in Fig.~\ref{fig_2}. In all cases $K(t)$ shows two distinct exponentially decreasing regimes with time constants $T_{1}$ and $T_{2}$ separated by a crossover region around $t_{\times}$. For fixed $N_{p}$, $T_{1}$ decreases monotonically as $F$ increases. The $t_{\times}$ also decreases as a function of $F$ leading to a faster overall degradation of the fiber. The situation is similar when $N_{p}$ is increased.

\begin{figure}
\begin{center}
\includegraphics[width=8cm]{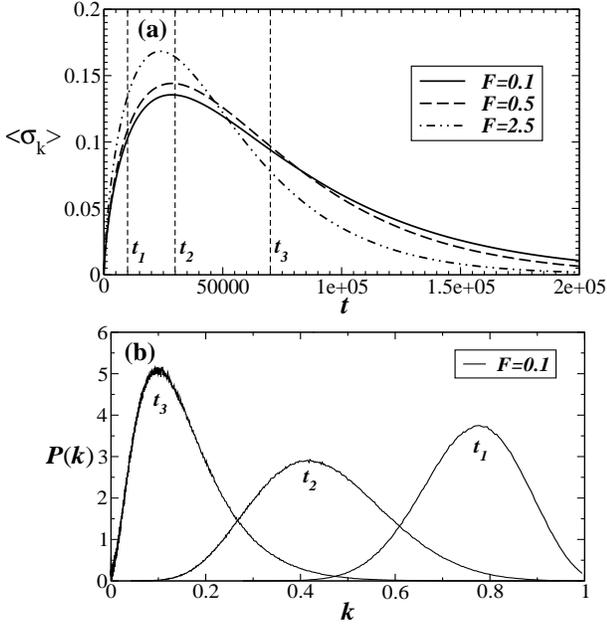}
\caption{$(a)$ The standard deviation $\langle \sigma_{k} \rangle$ of local spring constants $k$ as a function of time for three values of $F=0.1,0.5,2.5$ and $N_{p}=512$. In $(b)$, we plot the distribution of spring constants $P(k)$ at time points $t_{i}$ ($i=1,2,3$) as indicated in $(a)$ for $F=0.1$. In both graphs, we use the parameters $\Delta p=0.20$ and $p_{\text {off}}=0.5$.}
\label{fig_3}
\end{center}
\end{figure}

To characterize the changes in the microscopic properties of the fiber, we calculate the standard deviation $\langle \sigma_{k}\rangle$ of all spring constants at a fixed time and average them over all runs in Fig.~\ref{fig_3}$a$. Initially, for $t \ll t_{1}$,  $\langle \sigma_{k}\rangle$ increases quickly, which is not influenced by $F$. When $t\approx 2000$, $F$ starts to affect the binding process according to Eq.~(\ref{eq_1}) and hence $\langle \sigma_{k}\rangle$ increases faster for higher $F$. When $t$ is around $t_{2}$, $\langle \sigma_{k}\rangle$ reaches its maximun followed by a slow decay. The behavior can be confirmed by looking at the spring constant distributions $P(k)$ in Fig.~\ref{fig_3}$b$. The width of $P(k)$ has a maximum at $t_{2}$. For $t < t_{2} $ and $t>t_{2}$, $P(k)$ becomes narrower. Also, the peak of $P(k)$ decreases with increasing time. Similar behavior is observed for higher $N_{p}$; however, the maximum $\langle \sigma_{k}\rangle$ decreases when $N_{p}$ increases (data not shown).

To gain insight into the exponential decay of fiber stiffness, we carry out a simple mean field calculation. The local $k$ at time $t$ depends on the number of times the spring has been visited. We define $t_{i}$ as the time corresponding to the $i$-th unbinding event along the entire chain. Thus, $K$ at time $t_{i}$, can be written as

\begin{equation}
\begin{split}
 K(t_{i}) &=\frac{1}{\sum_{j=1}^{N_{s}} \frac{1}{k_{j}}} \simeq\frac{\langle k \rangle} {N_{s}} \\
\end{split}
\label{eq_2}
\end{equation}
assuming $\langle 1/k \rangle \simeq 1/ \langle k \rangle$. Notice that $K$ remains constant for $t_{i}\leqslant t < t_{i+1}$. At time $t_{i+1}$, an unbinding event occurs at spring $m$ and the corresponding $k_{m}$ is reduced to $\gamma k_{m}$. The new value of $K(t_{i+1})$ is

\begin{equation}
\begin{split}
 K(t_{i+1})  &=\frac{1}{\frac{1}{\gamma k_{m}} + \sum_{j\neq m}^{N_{s}} \frac{1}{k_{j}}} \simeq \frac{\langle k \rangle}{\left(\frac{1-\gamma}{\gamma}\right)+N_{s}} \\
\end{split}
\label{eq_3}
\end{equation}
where we also assume that $k_{m} \simeq \langle k \rangle$. Thus, from Eqs.~(\ref{eq_2}) and (\ref{eq_3}), the change $\Delta K= K(t_{i+1})-K(t_{i})$ in the total stiffness is written as

\begin{equation}
\begin{split}
\Delta K   &= K(t_{i+1})-K(t_{i})=-K\left[\frac{1}{1+\left(\frac{\gamma}{1-\gamma}\right)N_{s}}\right]. \\
\end{split}
\label{eq_4}
\end{equation}

\begin{figure}
 \begin{center}
  \includegraphics[width=0.95\columnwidth]{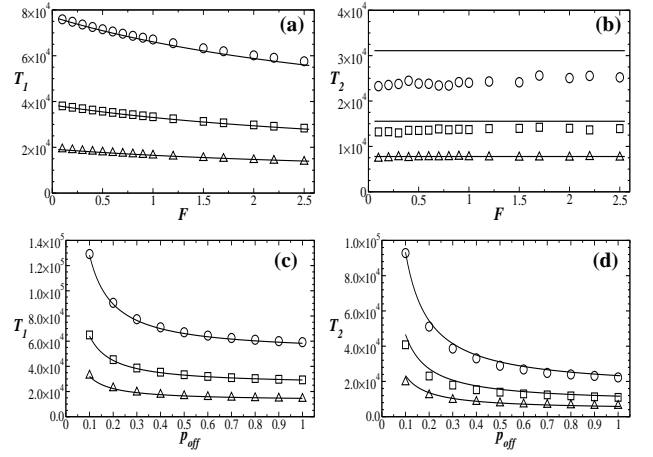}
\caption{Comparison of the mean field calculations and the numerical simulations. Time constants $T_{1}$ and $T_{2}$ are plotted as functions of $F$ and $p_{\text {off}}$ for $N_{p}=256$ (circles), $N_{p}=512$ (squares) and $N_{p}=1024$ (triangles). The symbols corespond to the numerical simulations and the solid lines are obtained from Eq.~(\ref{eq_8}). In $(a)$ and $(b)$, $T_{1}$ and $T_{2}$ as a function of $F$ for a constant $p_{\text {off}}=0.5$. In $(c)$ and $(d)$, $T_{1}$ and $T_{2}$ as a function of $p_{\text {off}}$ for $F=1.0$. In all graphs we used $\Delta p=0.20$.}
\label{fig_4}
\end{center}
\end{figure}

Next, we consider the average waiting time $\langle \tau \rangle$ between two unbinding events. During digestion, the number of particles $n_B$ that remains bound on the fiber changes, but $n_B$ is related to the number of free particles $n_F$ since $n_{B}+n_{F}=N_{p}$. The rate of change of $n_{B}$ is the difference between the average binding rate $p_{\text{on}}~n_{F}$ and the average unbinding rate $p_{\text{off}}~n_{B}$ which can be expressed as $~dn_{B}/dt=p_{\text {on}}~n_{F}-p_{\text {off}}~n_{B}$. Assuming that $n_{B}$ has reached a steady state $~dn_{B}/dt=0$, we obtain $n_{B}=p_{\text {on}} N_{p}/ \left (p_{\text {on}}+p_{\text {off}} \right )$.

Since the unbinding probability is per unit time, $\langle \tau \rangle$ can now be expressed as
\begin{equation}
 \langle \tau \rangle =\frac{1}{n_{B} p_{\text {off}}}= \frac{1}{N_{p}}\left(\frac{1}{p_{\text {on}}}+\frac{1}{p_{\text {off}}}\right)
\label{eq_5}
\end{equation}

Finally, we can establish a link between the two processes involved in enzymatic digestion. From Eq.~(\ref{eq_4}), $K$ is reduced by $\Delta K$ during the interval $\Delta t= t_{i+1}-t_{i}$. We thus approximate the derivative of $K$ by the discrete change $\Delta K$ during the interval $\Delta t \simeq \langle \tau \rangle$ as

 \begin{equation}
\newcommand{\D}{\displaystyle}
\normalsize
\begin{split}
 \frac{dK}{dt} & \simeq\frac{\Delta K}{\langle \tau \rangle} \simeq -\frac{K}{N_{s}} \left[ \left( \frac{1-\gamma}{\gamma} \right) \left( \frac{N_{p}}{\frac{1}{p_{\text {on}}}+\frac{1}{p_{\text {off}}}} \right) \right] \\
\end{split}
\label{eq_6}
\end{equation}
 
This equation can be solved assuming that $p_{\text {on}}$ is approximately constant during one time step. The result is given by $K(t)=e^{-t/T}$, where
\begin{equation}
T=\frac{\gamma}{1-\gamma} \left[\frac{1}{p_{\text {on}}}+\frac{1}{p_{\text {off}}} \right] \left(\frac{N_{s}}{N_{p}}\right).
\label{eq_7}
\end{equation}
The above expression represents different aspects of the digestion process. The first term which involves $\gamma$, is related to the geometry of the fiber, the average number of molecules in parallel. The second term is related to the specific enzyme activity at the microscopic level, the binding and unbinding processes. The third term is essentially the enzyme concentration that is an external control parameter. Note that the time constant $T$ is symmetric in the binding and unbinding probabilities. Thus, if we assume it is the unbinding process that depends on the external force, the results for the stiffness degradation will be identical.

We next analyze the asymptotic limits of Eq.~(\ref{eq_7}). We summarize the results as follows
\begin{equation}
T_{i}=
\begin{cases}
T_{1}=\frac{\gamma}{(1-\gamma)}\left[\frac{1}{\frac{1}{3}-\Delta p e^{-\lambda F}} + \frac{1}{p_{\text {off}}}\right]\left(\frac{N_{s}}{N_{p}}\right)&  \text{if $k\approx k_{0}$}\\
\\
T_{2}=\frac{\gamma}{(1-\gamma)}\left[3+\frac{1}{p_{\text {off}}}\right]\left(\frac{N_{s}}{N_{p}}\right)& \text{if $k\ll k_{0}$}.

\end{cases}
\label{eq_8}
\end{equation}

To compare the results of the numerical simulations with the analytical calculation in the asymptotic limits, we analyze the time course of stiffness by calculating the time constants in Fig.~\ref{fig_2} for the two different regimes of exponential behavior. We obtain exponential fits in non-overlapping windows with fixed size $\Delta t = 5000$ along the stiffness curve. The $T_{1}$ and $T_{2}$ are then extracted as a function of $F$ and $p_{\text {off}}$ in the regions $K(t)\approx 1$ and $K(t)\ll 1$, respectively. The results are compared to those obtained from Eq.~(\ref{eq_8}) in Fig.~\ref{fig_4}. Generally, the results confirm the agreement of $T_{1}$ and $T_{2}$ measured in the simulations and in the analytical calculations as a function of $F$ and $p_{\text {off}}$. The $T_{1}$ decreases with increasing $F$ while it diverges when $p_{\text {off}}$ becomes less than $1/3$. Notice also that for fixed values of $F$ and $p_{\text {off}}$, $T_{1}$ decreases with increasing enzyme concentration. Interestingly, $F$ has little effect on $T_{2}$. This can also be seen in Fig.~\ref{fig_2} which shows that after the crossover region all curves follow the same exponential decay for different $F$ at a fixed value of $N_{p}$, but decays slower as $N_{p}$ decreases. The difference in $T_{2}$ between analytical and numerical calculations for low $N_{p}$ is because $K$, in this limit, takes significantly more time to reaches the second exponential regime. To confirm this, we run an additional simulation for $N_{p}=256$ for $3\times 10^{5}$ time steps and find that $T_{2}$ increases by about $15\%$, approaching the analytical results. Also for a fixed $F$ and $p_{\text {off}}$, the $T_{1}$ increases as $\Delta p$ increases and  $T_{2}$ remains approximately constant as a function of $\Delta p$ (data not shown).

In summary, we have presented a model for the enzymatic digestion of an elastic fiber under tension. We have shown that the total stiffness decreases exponentially with two different regimes separated by a crossover region. While the first exponential regime has been found experimentally \cite{Rajiv_07}, to our knowledge, the second one has not been measured and remains a prediction of the model. Each regime can be associated with an average value of the stiffness along the fiber during the particle diffusion process. In the first regime, the stiffness is dominated by the average local initial stiffness values whereas in the second regime, the local stiffness has decreased significantly and almost uniformly throughout the fiber. In the crossover region, the stiffness is controlled by a wide distribution of local stiffness values. The time constant $T_{1}$ displays a strong dependence on both $F$ and $N_{p}$. This result is in agreement with the experimental results \cite{Rajiv_07} where it was reported they show that application of static mechanical forces accelerates the digestion-induced breakdown of ECM sheets. Also, we have performed analytical calculations that confirm the presence of two different regimes. These calculations also show how the time constant can be partitioned into geometrical, chemical and externally controllable factors. Furthermore, in the first regime, we expect that the fiber does not reach the failure limit but after the crossover the decrease in the stiffness is more intense and failure is likely to occur. These results can help better understand natural growth and maintenance of the ECM as well as diseases in which enzyme concentrations and/or mechanical forces become abnormal.

This work is supported by the Brazilian agency CNPq and NIH HL59215 and HL090757.


\begin{thebibliography}{99}

\bibitem{Ghajar_08} C. M. Ghajar, S. C. George and  A. J. Putnam, Crit Rev Eukaryot Gene Expr. { \bf 18}, 251-78 (2008).

\bibitem{Barnes_00} P. J. Barnes, N. Engl. J. Med. {\bf 343}, 269-280 (2000).

\bibitem{Rajiv_07} R. Jesudason, L. Black, A. Majumdar, P. Stone and B. Suki, J. Appl. Physiol. {\bf 103}, 803-811 (2007).

\bibitem{Abete_04} T. Abete, A. de Candia, D. Lairez and A. Coniglio, Phys. Rev. Lett. {\bf
93}, 228301 (2004).

\bibitem{Dahirel_09} V. Dahirel, F. Paillusson, M. Jardat, M. Barbi and J.-M. Victor, Phys. Rev. Lett. {\bf
102}, 228101 (2009).

\bibitem{Schmit_09} J. D. Schmit, E. Kamber and J. Kondev, Phys. Rev. Lett. {\bf 102}, 218302 (2009).

\bibitem{Berry_02} H. Berry, Biophys. J. {\bf 83}, 1891 (2002).

\bibitem{Michael_05} M. I. Minine and J. Haugh, J. Chem. Phys. {\bf 123}, 074908 (2005).

\bibitem{Gennes_82} P. G. Gennes, J. Chem. Phys. {\bf 76}, 3316 (1982).

\bibitem{Kantor_84} Y. Kantor and I. Webman, Phys. Rev. Lett. {\bf 52}, 1891 (1984).

\bibitem{Bouchiat_98} C. Bouchiat and M. M\'{e}zard, Phys. Rev. Lett. {\bf 80}, 1556 (1998).
\end{thebibliography}
\end{document}